\documentclass[conference]{IEEEtran}

\IEEEoverridecommandlockouts
\usepackage{cite}
\usepackage{amsmath,amssymb,amsfonts}
\usepackage{algorithmic}
\usepackage{graphicx}
\usepackage{textcomp}
\usepackage[table]{xcolor}  % for colors in tables
\usepackage{booktabs, makecell, booktabs,multirow} % for figures and tables subfig, 
\usepackage{url} % to insert url
\usepackage{hyperref}
\usepackage{subcaption}

\def\BibTeX{{\rm B\kern-.05em{\sc i\kern-.025em b}\kern-.08em
    T\kern-.1667em\lower.7ex\hbox{E}\kern-.125emX}}
    
\newcommand{\etal}{\textit{et al}.\@ }
\newcommand{\ie}{\textit{i.e.},\@ }

\newcolumntype{L}[1]{>{\raggedright\let\newline\\\arraybackslash\hspace{0pt}}m{#1}}
\newcolumntype{C}[1]{>{\centering\let\newline\\\arraybackslash\hspace{0pt}}m{#1}}
\newcolumntype{R}[1]{>{\raggedleft\let\newline\\\arraybackslash\hspace{0pt}}m{#1}}

\usepackage{stackengine}

\definecolor{emerald}{rgb}{0.31, 0.78, 0.47}
\definecolor{OliveGreen}{rgb}{0,0.7,0}
\definecolor{caribbeangreen}{rgb}{0.0, 0.8, 0.6}
\definecolor{ao(english)}{rgb}{0.0, 0.5, 0.0}
\definecolor{kellygreen}{rgb}{0.3, 0.73, 0.09}
\title{Synthesized Speech Detection Using Convolutional\\Transformer-Based Spectrogram Analysis
}

\author{
    \IEEEauthorblockN{\textbf{Emily R. Bartusiak, Edward J. Delp}}
    \vspace{.25cm}
    \IEEEauthorblockA{Video and Image Processing Lab (VIPER) \\
    School of Electrical and Computer Engineering\\
    Purdue University\\
    West Lafayette, IN, USA}
}

\begin{document}
\maketitle

%%%%%%%%%%%%%%%%%%%%%%%%%%%%%%%%%
% Abstract
%%%%%%%%%%%%%%%%%%%%%%%%%%%%%%%%%
\begin{abstract}
Synthesized speech is common today due to the prevalence of virtual assistants, easy-to-use tools for generating and modifying speech signals, and remote work practices. 
Synthesized speech can also be used for nefarious  purposes, including creating a purported speech signal and attributing it to someone who did not speak the content of the signal.
We need methods to detect if a speech signal is synthesized.
In this paper, we analyze speech signals in the form of spectrograms with a Compact Convolutional Transformer (CCT) for synthesized speech detection.
A CCT utilizes a convolutional layer that introduces inductive biases and shared weights into a network, allowing a transformer architecture to perform well with fewer data samples used for training.
The CCT uses an attention mechanism to incorporate information from all parts of a signal under analysis.
Trained on both genuine human voice signals and synthesized human voice signals, we demonstrate that our CCT approach successfully differentiates between genuine and synthesized speech signals.
\end{abstract}

%%%%%%%%%%%%%%%%%%%%%%%%%%%%%%%%%
% Keywords
%%%%%%%%%%%%%%%%%%%%%%%%%%%%%%%%%

%%%%%%%%%%%%%%%%%%%%%%%%%%%%%%%%%%%%%%
% Paper Content
%%%%%%%%%%%%%%%%%%%%%%%%%%%%%%%%%%%%%%
\section{Introduction}\label{part-1-intro}

We hear more synthesized voices in our daily lives than ever before.
Voice assistants, such as Apple's Siri, Amazon's Alexa, Microsoft's Cortana, and Google's Assistant, use synthesized human speech to communicate with us in our homes~\cite{hoy_2018, malodia_2021}.
Virtual assistants answer customer service phone numbers and create synthetic voices as they assist us. 
Applications use text-to-speak (TTS) methods to generate audio signals to speak text messages for users with poor vision. 
Social media platforms offer tools to generate speech that sounds like a specific person, such as a friend, an actor, or a politician.
Although all of these features can be used for innocuous purposes, they can also easily be used to create authentic-sounding speech for more malicious ambitions.

Attackers may create a purported speech signal and attribute it to someone who never delivered that message.
In 2021, Goldman Sachs stopped a \$40 million investment in a company when employees realized they were meeting with an impersonator using synthesized speech on a conference call~\cite{smith_2021}.
Although standalone synthesized speech replicating a specific target's voice can do a lot of damage on its own, its impact can be even greater when the synthesized speech is paired with other data modalities.
When manipulated video accompanies synthetic speech signals, such as in deepfakes, the potential to influence public opinion and current events is even higher~\cite{rossler_2019, toews_2020}.
Because many easy-to-use tools exist for modifying multimedia with high quality, the quantity of manipulated media increases exponentially~\cite{patrini_2019}.
We need methods to detect if speech signals are synthesized or genuine.

%%%%%%%%%%%%%%%%%%%%%%%%%%%%%%%%%%%%%%%%%%%%%%%%%%%%%%%%%%%%%%%%%%%%%%%%%%%
\begin{figure}[t]
    \centering
    \includegraphics[width=4cm]{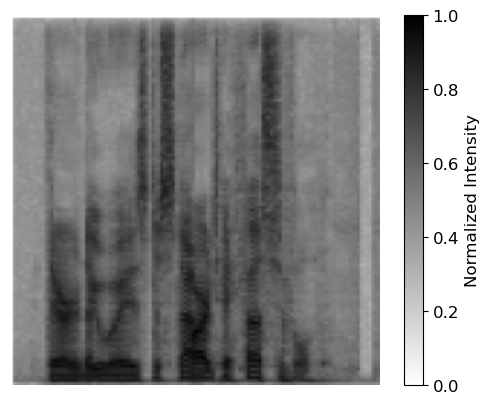}
    \caption{Spectrogram of a synthesized speech signal.}
    \label{fig:input}
\end{figure}

In this paper, we present a method to detect synthesized speech
from spectrograms. 
Spectrograms have been used for a variety of audio tasks to identify events, transfer a signal into a new style, and detect emotion~\cite{flanagan_1972, verma_2017, dennis_2011, stolar_2018, pras_2015, pras_2015_2}. 
They are effective visualizations of speech signals because they show the relationship between time, frequency, and intensity of an audio signal. 
Spectrograms are constructed using the Fast Fourier Transform (FFT), a variant of the Discrete Fourier Transform (DFT)~\cite{flanagan_1972}.
The FFT divides speech signals into shorter temporal subsequences (sometimes known as frames). 
The DFT of each temporal segment is computed to obtain the DFT frequency coefficients for each temporal segment. 
The magnitudes of the DFT coefficients are aligned side-by-side to create a spectrogram. 
A spectrogram of a speech signal used in our approach is shown in Figure~\ref{fig:input}.
In this paper, we treat spectrograms as images and use a convolutional transformer-based approach to determine if the speech signals shown in the spectrograms are synthesized.

%%%%%%%%%%%%%%%%%%%%%%%%%%%%%%%%%%%%%%%%%%%%%%%%%%%%%%%%%%%%%%%%%%%%%%%%%%%
\begin{figure*}[ht]
    \centering
    
    \includegraphics[width=.9\textwidth]{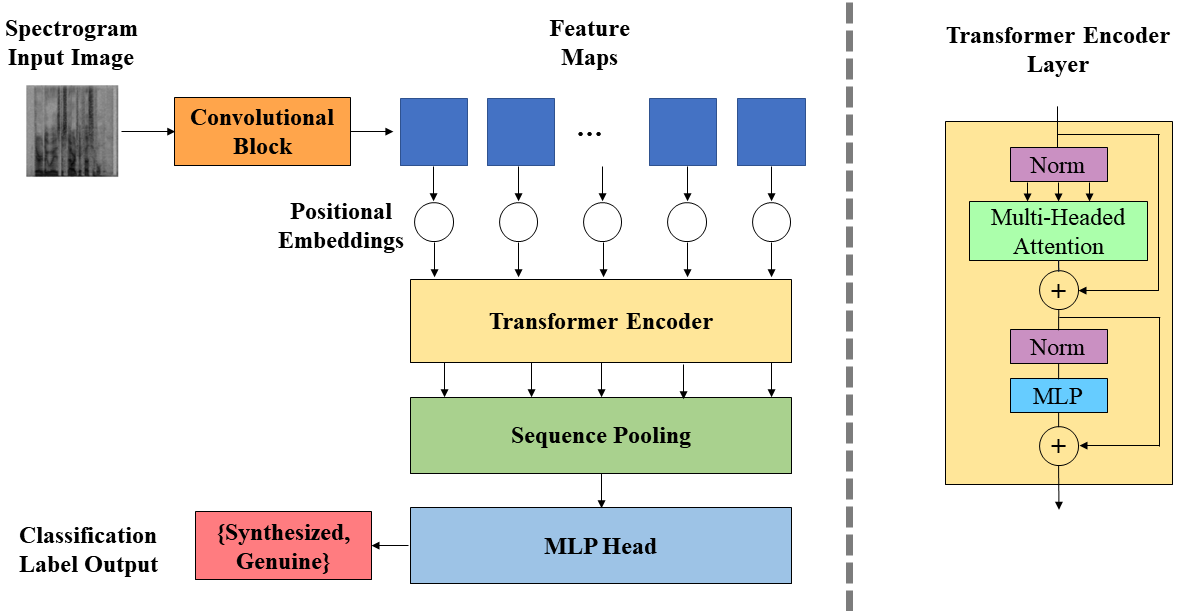}
%   \subfloat{\includegraphics[width=.9\textwidth]{figures/cct.png}}
    \caption{Block diagram of the proposed approach: the Compact Convolutional Transformer (CCT).}
    \label{fig:CCT}
\end{figure*}
%%%%%%%%%%%%%%%%%%%%%%%%%%%%%%%%%%%%%%%%%%%%%%%%%%%%%%%%%%%%%%%%%%%%%%%%%%%

\section{Related Work}

In order to detect synthetic speech, various machine learning methods convert speech signals into
different representations.
Chen \etal use a Multilayer Perceptron Network (MLP), ResNet-based Convolutional Neural Network (CNN), Long Short-Term Memory network (LSTM), Gated Recurrent Unit network (GRU), and Recurrent Neural Network (RNN) to detect spoofed speech~\cite{chen_2017, chen_2018}. 
For these methods, the speech waveforms are converted to sequences of coefficients -- specifically Constant Q Cepstral Coefficients (CQCCs)~\cite{bogert} or Mel Frequency Cepstral Coefficients (MFCCs)~\cite{mel, purves_2001, mfccs} -- and analyzed by the various neural networks. 
Chintha \etal use a CNN-LSTM model to analyze speech signals directly, rather than converting them to different coefficient representations~\cite{chintha_2020}.
The authors also explore working with log-melspectrograms, which are spectrograms in which the frequency domain content is mapped to the mel scale~\cite{mel}.
The log-melspectrograms are analyzed with a CNN to detect synthesized and genuine audio.
For our approach, we utilize spectrograms because many methods striving to perform a variety of tasks achieve success with signals represented as spectrograms~\cite{flanagan_1972, verma_2017, dennis_2011, stolar_2018, pras_2015, pras_2015_2}.

Although prior work relies heavily on Convolutional Neural Networks (CNNs), recent developments in deep learning indicate that convolutions may not be necessary to effectively analyze images~\cite{dosovitskiy_2021, tolstikhin_2021}. These methods succeed in image classification tasks without inductive biases provided by convolutions. Inspired by the success of attention mechanisms in Natural Language Processing (NLP)~\cite{vaswani_2017}, Vision Transformer (ViT) analyzes patches of an image with an attention mechanism for image classification tasks~\cite{dosovitskiy_2021}. Hassani \etal adopt concepts from CNNs and ViT to create a Compact Convolutional Transformer (CCT)~\cite{hassani_2021}. 
CCT leverages the inductive biases and efficiencies of parameter-sharing that convolutions provide to succeed at machine learning tasks with smaller-sized datasets compared to the datasets used with ViT.
It also leverages the attention mechanism of transformers to capture long-range dependencies in images.
CCT combines the power of convolutions with the power of transformers.
We utilize a CCT trained on spectrograms showing genuine and synthesized speech signals to identify synthesized speech.
\section{Proposed Method}

Figure~\ref{fig:CCT} shows an overview of our approach, known as the Compact Convolutional Transformer (CCT).
The CCT uses a standard transformer encoder, as used in ~\cite{vaswani_2017, dosovitskiy_2021}.
However, CCT introduces two new features -- a convolutional image encoding block and a sequence pooling layer -- that replace operations in standard transformer approaches.
The CCT first uses a convolutional block (\ie a series of convolutional layers) to embed an input image into a latent space. 
In our experiments, we utilize two convolutional layers with a kernel of size 3x3, a ReLU activation function, and max pooling.
The first and second convolutional layers produce sets of 64 and 128 feature maps, respectively.
We use this convolutional block instead of the standard transformer practice of dividing input images into non-overlapping patches, which contain only local image information and fail to preserve information at patch boundaries.
The feature maps contain aggregate information from all regions of an image, so they are more salient inputs to the transformer encoder.
Because they result from convolution operations, they also introduce inductive biases to the network.
This enables the transformer to train more efficiently, which is highly important on smaller-sized datasets.
Next, we row concatenate each of the 128 2D feature maps (sized 32x32) into a vector of length 1024, creating the tokens analyzed by the transformer encoder.
We use positional embedding (a standard practice in transformers) for each token so that the transformer understands how they relate spatially~\cite{dosovitskiy_2021, bert_devlin_2019}. 

Next, CCT analyzes the tokens with the transformer encoder, which consists of two transformer encoder layers.
Each transformer encoder layer contains the multi-headed attention mechanism that captures long-range dependencies between different parts of the input. 
The transformer encoder layers are modeled after typical attention-based layers~\cite{dosovitskiy_2021, vaswani_2017}.
Then, sequence pooling occurs on the outputs of the transformer encoder.
The pooling operation smooths the sequence of outputs so that the MLP Head can correctly detect whether the speech signal under analysis is \textit{synthesized} or \textit{genuine}. 
Sequence pooling also eliminates the need for an extra token (\ie a classification token) that other transformers use~\cite{dosovitskiy_2021, bert_devlin_2019}.
With sequence pooling, the model no longer needs to track the classification token throughout its layers.

%%%%%%%%%%%%%%%%%%%%%%%%%%%%%%%%%%%%%%%%%%%%%%%%%%%%%%%%%%%%%%%%%%%%%%%%%%%
\begin{table}[ht]
{\rowcolors{3}{white!10}{kellygreen!10}
\begin{center}
    \caption{Dataset used in our experiments.}
    {\renewcommand{\arraystretch}{1.5} %<- modify value to suit your needs
    \begin{tabular}{ |C{1.6cm}|C{1.6cm}|C{1.6cm}|C{1.6cm}|  }
    \hline
    \rowcolor{kellygreen!40} \multicolumn{4}{|c|}{\textbf{ASVspoof2019 Dataset}} \\
    \hline
    \rowcolor{kellygreen!15!}
    Subset & Synthesized \newline Audio Tracks & Genuine \newline Audio Tracks & Total \newline Audio Tracks \\
    \hline
    Training    &  22,800 &  2,580 &  25,380 \\
    Validation  &  22,296 &  2,548 &  24,844 \\
    Testing     &  63,882 &  7,355 &  71,237 \\
    Total       & 108,978 & 12,483 & 121,461 \\
    \hline
    \end{tabular}
    }
    \label{tab:dataset}
\end{center}
}
\end{table}
%%%%%%%%%%%%%%%%%%%%%%%%%%%%%%%%%%%%%%%%%%%%%%%%%%%%%%%%%%%%%%%%%%%%%%%%%%%

\section{Experimental Results}

We utilize the ASVspoof2019 dataset~\cite{asvdata_2019} in our experiments. 
The dataset -- introduced in ASVspoof2019: Automatic Speaker Verification Spoofing and Countermeasures Challenge~\cite{asvplan_2019} -- contains \textit{genuine} speech signals spoken by humans as well as \textit{synthesized} speech signals.
The synthesized speech signals were generated with neural acoustic models and deep learning methods, including LSTMs~\cite{hochreiter_1997} and Generative Adversarial Networks (GANs)~\cite{goodfellow_2014}. 
The ASVspoof2019 dataset is heavily imbalanced, with significantly more synthesized speech signals than genuine speech signals.
We utilize the official dataset split according to the challenge for training, validating, and testing our approach.
Table~\ref{tab:dataset} summarizes the details of the dataset. 

We convert speech waveforms from the dataset into spectrograms by following a similar procedure as described in~\cite{bartusiak_2021}.
More specifically, we use the Fast Fourier Transform (FFT) to compute Fourier coefficients of signals in our dataset.
The FFT operates on blocks of the signals consisting of 512 sampled points with 128 points of overlap between consecutive blocks.
Then, the Fourier coefficients are converted to decibels and organized in 2D arrays to construct the spectrograms.
We represent each spectrogram with a matrix of 128x128 values. 
Note that these spectrograms are larger than those used in our previous approach~\cite{bartusiak_2021}.
The larger spectrograms have higher resolution, which preserves more details of speech signals for the synthesized speech detector.
Next, we perform min-max normalization on the intensity values, mapping the spectrogram intensities to the range of values [0,1].
Normalized values enable machine learning models to learn more quickly because they are forced to focus on relative rather than absolute differences in input values. 
Figure~\ref{fig:input} shows an example of a grayscale, normalized spectrogram that is analyzed by the CCT.

To validate our approach, we compare it against several other methods.
First, we establish three baseline methods: Baseline-Minority, Baseline-Majority, and Baseline-Prior.
Baseline-Minority is a classifier that only predicts that speech signals belong to the minority class -- in this case, the \textit{genuine} category.
Baseline-Majority is a classifier that does the opposite.
It only predicts that speech signals belong to the \textit{synthesized} class.
Considering the significant class imbalance in the dataset, we expect Baseline-Minority to have the worst performance of all methods.
Meanwhile, Baseline-Majority establishes a threshold above which a classifier actually performs well.
Baseline-Prior is the final baseline classifier. 
It randomly assigns a label to a signal under analysis according to the known distribution of \textit{genuine} vs. \textit{synthesized} samples in the training data split.
In addition to these baselines, we investigate the effectiveness of a K-Nearest Neighbors classifier~\cite{silverman_1989, cover_1967}, a Support Vector Machine (SVM)~\cite{cortes_1995}, and logistic regression~\cite{Bishop_2006} on this task.
These methods operate on row concatenated versions of the 128x128-sized spectrograms (\ie vectors of length 16,384). 
Finally, we compare our results to our previous synthesized speech detection method that utilizes a CNN~\cite{bartusiak_2021}.
We report the performance of the CNN on the smaller-sized spectrograms (dimensions 50x34 pixels) as well as the the new, larger-sized spectrograms (dimensions 128x128).

%%%%%%%%%%%%%%%%%%%%%%%%%%%%%%%%%%%%%%%%%%%%%%%%%%%%%%%%%%%%%%%%%%%%%%%%%%%
\begin{figure*}[ht]
    \centering
    \includegraphics[width=.9\textwidth]{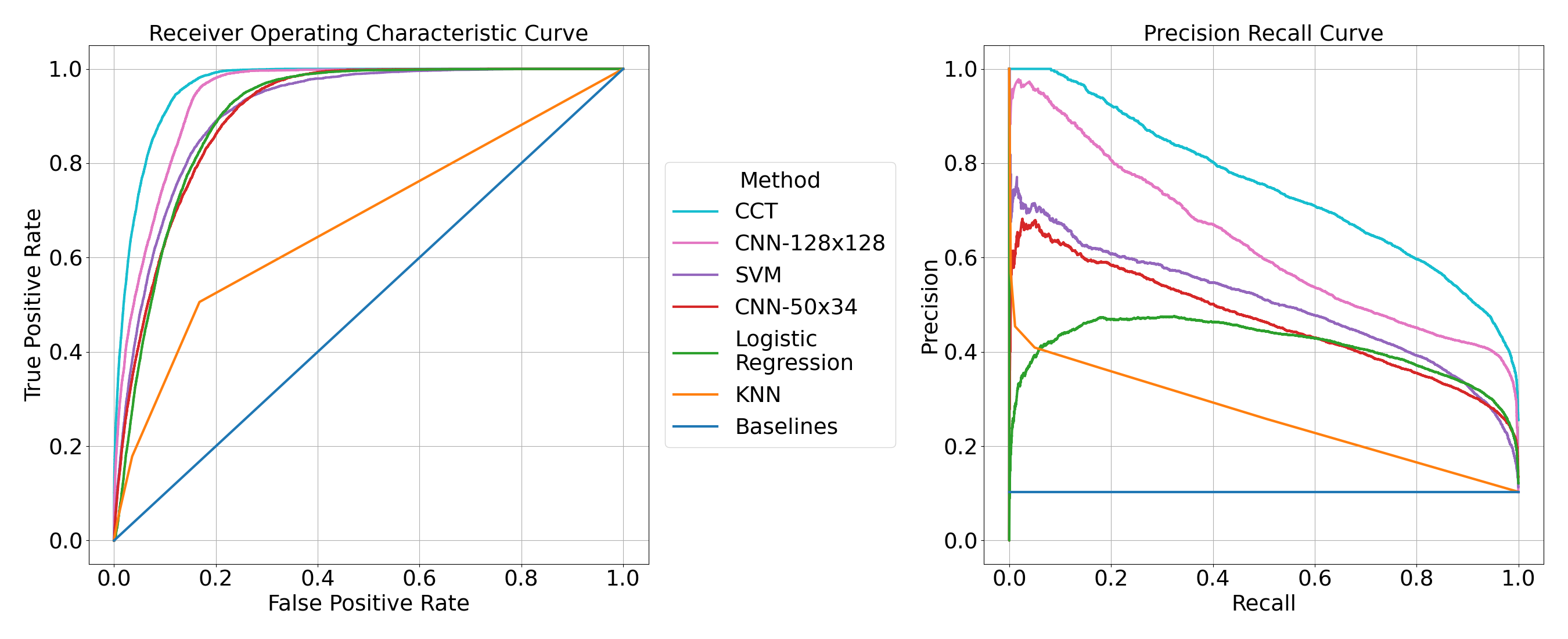}
    \caption{The ROC and PR of all methods examined. The three baseline methods are all represented by the ``baselines" method.}
    \label{fig:curves}
\end{figure*}

%%%%%%%%%%%%%%%%%%%%%%%%%%%%%%%%%%%%%%%%%%%%%%%%%%%%%%%%%%%%%%%%%%%%%%%%%%%

%%%%%%%%%%%%%%%%%%%%%%%%%%%%%%%%%%%%%%%%%%%%%%%%%%%%%%%%%%%%%%%%%%%%%%%%%%%
% \begin{table*}[ht]
%     \begin{center}
%     \caption{\textbf{Macro Results.} This table indicates the performances of the baseline random method and our proposed method.}
%     \resizebox{1.5\columnwidth}{!}{
%         \begin{tabular}{cccccc}
% 	        \toprule
%             \textbf{Method} &  \textbf{Accuracy} & \textbf{Precision} & \textbf{Recall} & \textbf{F-1} & \textbf{Balanced Accuracy}\\
%             \midrule
%             Baseline (Majority)     & 89.68\% & 44.84\% & 50.00\% & 47.28\% & 50.00\%\\
%             Baseline (Minority)     & 10.32\% & 5.16\% & 50.00\% & 9.36\% & 50.00\%\\
%             Baseline (Prior)        & 81.54\% & 49.81\% & 49.81\% & 49.81\% & 49.81\%\\
%             CNN                     & 85.59\% & 67.63\% & 79.26\% & 70.95\% & 79.26\%\\
%             CCT                     & \textbf{92.13\%} & \textbf{78.15\%} & \textbf{87.78\%} & \textbf{81.92\%} & \textbf{87.78\%}\\
% 	        \bottomrule
%         \end{tabular}}
%         \label{tab:results-macro}
%   \end{center}	
% \end{table*}
%%%%%%%%%%%%%%%%%%%%%%%%%%%%%%%%%%%%%%%%%%%%%%%%%%%%%%%%%%%%%%%%%%%%%%%%%%%

%%%%%%%%%%%%%%%%%%%%%%%%%%%%%%%%%%%%%%%%%%%%%%%%%%%%%%%%%%%%%%%%%%%%%%%%%%%
\begin{table*}[ht]
    \begin{center}
    \caption{Results of all methods. ROC AUC and PR AUC represent area under the curve of the receiver operating characteristic and precision recall curves, respectively.}
    \resizebox{2\columnwidth}{!}{
        {\renewcommand{\arraystretch}{1.1} %<- modify value to suit your needs
        \begin{tabular}{lccccccc}
	        \toprule
            \textbf{Method} &  \textbf{Accuracy} & \textbf{Precision} & \textbf{Recall} & \textbf{F-1} & \centering \textbf{Balanced  Accuracy} & \textbf{ROC AUC} & \textbf{PR AUC}\\
            \midrule
            Baseline-Minority     & 10.32\% & 1.07\% & 10.32\% & 1.93\% & 50.00\% & 0.5000 & 0.1032\\
            Baseline-Prior        & 81.59\% & 81.46\% & 81.59\% & 81.52\% & 49.94\% & 0.4994 & 0.1032\\
            Baseline-Majority     & 89.68\% & 80.42\% & 89.68\% & 84.79\% & 50.00\% & 0.5000 & 0.1032\\
            KNN                   & 89.45\% & 84.99\% & 89.45\% & 85.57\% & 52.08\% & 0.6751 & 0.2643\\
            Logistic Regression   & 80.60\% & 91.76\% & 80.60\% & 83.98\% & 84.41\% & 0.9041 & 0.4101\\
            CNN-50x34             & 85.59\% & 90.50\% & 85.59\% & 87.35\% & 79.26\% & 0.9052 & 0.4649\\
            SVM                   & 89.93\% & 90.94\% & 89.93\% & 85.25\% & 50.47\% & 0.9113 & 0.5023\\
            CNN-128x128           & 85.27\% & 93.21\% & 85.27\% & 87.60\% & \textbf{89.22}\% & 0.9416 & 0.6278\\
            CCT                   & \textbf{92.13\%} & \textbf{93.79\%} & \textbf{92.13\%} & \textbf{92.70\%} & 87.78\%  & \textbf{0.9646} & \textbf{0.7501}\\
	        \bottomrule
        \end{tabular}}
        }
        \label{tab:results-weighted}
  \end{center}	
\end{table*}
%%%%%%%%%%%%%%%%%%%%%%%%%%%%%%%%%%%%%%%%%%%%%%%%%%%%%%%%%%%%%%%%%%%%%%%%%%%

Table~\ref{tab:results-weighted} and Figure~\ref{fig:curves} show the results of all approaches. 
We report accuracy, weighted precision, weighted recall, weighted F-1, Balanced Accuracy, Receiver Operating Characteristic Area Under the Curve (ROC AUC), and Precision Recall Area Under the Curve (PR AUC)~\cite{tharwat_2021, davis_2006}. 
Weighted metrics are computed with a weighted average of each metric obtained on the two classes, where weights reflect the dataset class imbalance.
Results indicate that the CCT approach outperforms all other methods by a clear margin.
It achieves the highest metrics of all methods considered.
Although the CCT performs better than our previously proposed CNN, the CNN achieves the second highest performance when trained on spectrograms of size 128x128 pixels.
Overall, the two neural network approaches perform the best.

Results confirm that both larger input images and the new model contribute to better performance.
Comparing the results of CNN-50x34 and CNN-128x128, we observe that balanced accuracy, ROC AUC, and PR AUC increase when the CNN is trained and evaluated on larger, higher-resolution inputs. 
In this case, the CNN is presented with more detailed input images that allow it to better discriminate genuine and synthesized speech signals.
Comparing the results of CNN-128x128 and CCT, we observe that ROC AUC and PR AUC increase even more when an attention mechanism is used. 
The attention mechanism of the transformer determines the most important part of a spectrogram and focuses on that part of the image more so than the less discriminative regions, which aids in its detection capabilities.
However, transformers have historically required very large-scale datasets in order to learn properly, suffering from a lack of inductive biases that CNNs have.
Because we utilize convolutional layers, the CCT achieves a greater degree of shared weights, learns more efficiently, and leverages the inductive biases to achieve high success, even with fewer data samples from which to learn.

\section{Conclusion}
%%%%%%%%%%%%%%%%%%%%%%%%%%%%%%%%%%%%%%%%%%%%%%%%%%%%%%%%%%%%%%%%%%%%%%%%%%%
This paper improves upon our previous work~\cite{bartusiak_2021} to demonstrate the benefits of using higher-resolution spectrograms and an attention mechanism.
We demonstrate that a neural network that utilizes both convolution and transformer capabilities achieves high success in detecting synthesized speech. 
Convolution operations convert a spectrogram input image into feature maps that contain salient information for discriminating synthesized and genuine human speech.
It also enables a transformer to achieve high success with less data than transformers typically require.

Although this approach is promising, future work should consider more diverse speech features.
For example, the method should be validated on data of different audio formats, compression levels, sampling rates, and durations.
Since both the CNN and the CCT perform well, an ensemble of these two methods could be created and augmented with other neural networks.
Finally, a speech analysis method such as this could be paired with methods that analyze media's other data modalities. 
For example, our synthesized speech detector could analyze speech signals found in videos, while methods that analyze images and videos could analyze the visual content~\cite{mas_2020}~\cite{rossler_2019}~\cite{guera_2018}~\cite{bartusiak_2019}. 
A metadata analysis could strengthen this multi-modal approach even more~\cite{guera_2019}.

%%%%%%%%%%%%%%%%%%%%%%%%%%%%%%%%%
% Acknowledgements
%%%%%%%%%%%%%%%%%%%%%%%%%%%%%%%%%
\section*{Acknowledgment}
This material is based on research sponsored by DARPA and Air Force Research Laboratory (AFRL) under agreement number FA8750-16-2-0173. The U.S. Government is authorized to reproduce and distribute reprints for Governmental purposes notwithstanding any copyright notation thereon. The views and conclusions contained herein are those of the authors and should not be interpreted as necessarily representing the official policies or endorsements, either expressed or implied, of DARPA and AFRL or the U.S. Government.

Address all correspondence to Edward J. Delp, \\ \url{ace@ecn.purdue.edu}. 

%%%%%%%%%%%%%%%%%%%%%%%%%%%%%%%%%%
% Bibliography
%%%%%%%%%%%%%%%%%%%%%%%%%%%%%%%%%%
\bibliographystyle{IEEEtran}
\bibliography{refs}

\end{document}